# An investigation on the skewness patterns and fractal nature of research productivity distributions at field and discipline level[1]

*Giovanni Abramo*[*]
   Laboratory for Studies in Research Evaluation, Institute for System Analysis and Computer Science (IASI-CNR), National Research Council of Italy
      Via dei Taurini 19, 00185 Rome, Italy
      giovanni.abramo@uniroma2.it

*Ciriaco Andrea D'Angelo*
   Department of Engineering and Management, University of Rome 'Tor Vergata' and Laboratory for Studies in Research Evaluation, Institute for System Analysis and Computer Science (IASI-CNR)
      Via del Politecnico 1, 00133 Rome, Italy
      dangelo@dii.uniroma2.it

*Anastasiia Soldatenkova*
   Department of Engineering and Management, University of Rome 'Tor Vergata'
      Via del Politecnico 1, 00133 Rome, Italy
      anastasiia.soldatenkova@uniroma2.it

**Abstract**

The paper provides an empirical examination of how research productivity distributions differ across scientific fields and disciplines. Productivity is measured using the FSS indicator, which embeds both quantity and impact of output. The population studied consists of over 31,000 scientists in 180 fields (10 aggregate disciplines) of a national research system. The Characteristic Scores and Scale technique is used to investigate the distribution patterns for the different fields and disciplines. Research productivity distributions are found to be asymmetrical at the field level, although the degree of skewness varies substantially among the fields within the aggregate disciplines. We also examine whether the field productivity distributions show a fractal nature, which reveals an exception more than a rule. Differently, for the disciplines, the partitions of the distributions show skewed patterns that are highly similar.

**Keywords**
*Bibliometrics, research evaluation, FSS, Italy, CSS*

[*] *Corresponding author*

# 1. Introduction

A number of social phenomena do not show the common normal distributions. Classic examples include the cases of income, wealth and prices, for which most observations are concentrated towards the lower limit, and where distributions show strong skewness with long tail on the right, implying inequality.

Scientific activity is another social phenomenon whose main indicators are widely considered to be unequal in distribution. The literature provides empirical evidence on the subject, particularly through observation of two standard measures of researcher performance: numbers of publications produced and citations to the publications. Studies of skewness in the distribution of citations originate with Seglen (1992), and demonstrate that inequality in impact appears in various disciplines and fields and at different levels of aggregation (among recent works: Albarrán, Crespo, Ortuño, & Ruiz-Castillo, 2011; Franceschet, 2011; Chatterjee, Ghosh, & Chakrabarti, 2016; Ruiz-Castillo, 2012; Bornmann & Leydesdorff, in press).

Lotka (1926) originally wrote on the frequency distribution of number of publications; since then this metric has generally been considered to show research productivity. Although his study did not emphasize the concept of skewness, "Lotka's Law" has come to imply that most researchers have a small number of published papers. Later research on productivity distribution asymmetry has concentrated on verifying the law in different fields, using data on publication counts. In our view, the most comprehensive investigation into skewness of performance distribution across fields is the one by Ruiz-Castillo and Costas (2014). These authors studied the shape of productivity distributions as measured by number of articles and mean citation per publication. Their field of observation consisted of 17.2 million disambiguated world authors, whose Web of Science (WoS) indexed publications in the period 2003–2011 were classified into 30 broad scientific fields. The main finding is that the distributions are highly skewed and have similar patterns. The analyses for the population as a whole, and for the part above the first mean value, also revealed the fractal nature of the distributions – an issue which we will return to later in the paper. Ruiz-Castillo and Costas took the only approach possible when examining performance distributions at the world level, which is to begin from the WoS indexed publications grouped by field, and from these identify and disambiguate the authors. However, in this paper we exploit a distinctive feature of the Italian university system, which is that every professor is classified into one and only one research field. This allows us to start from the researchers rather than their publications. Consequently, we are able to examine classes of researchers, rather than the examining those who at a given time publish in the different fields. As we next explain, the implications are significant.

Our approach is to begin from the 370 fields (called "Scientific Disciplinary Sectors, SDSs) of the Italian research system, which in turn group the researchers under 14 disciplines ("University Disciplinary Areas", UDAs). Using a disambiguation algorithm developed by D'Angelo, Giuffrida & Abramo (2011), we then associate each professor with his/her WoS publications for the period under study. The approach offers immediate advantages. First, we can spot the unproductive researchers working in a particular field. Second, given that authors can publish in different fields, we are able to measure their real productivity, independent of how they diversify output among fields. To exemplify, in our approach, if a statistician publishes five works on statistical modelling and five on epidemiology, her performance by number of publications is 10.

Differently, using any approach based on field classification of output, her performance would only be five as a statistician, while she would also show a performance of five among physicians (which she is not). Furthermore, we use "Fractional Scientific Strength" (FSS) as the indicator of productivity. This indicator embeds both the number of publications and their relative impact (Abramo and D'Angelo, 2014), thus addressing the weaknesses of performance indicators that rely on number of publications alone, or on mean citations per publication. We have examined the problems of such indicators in two specific works, published in this same journal (Abramo and D'Angelo, 2016a; 2016b).

The literature provides very broad evidence of skewness in research productivity, whether measured by quantity or impact. Given that FSS embeds both, we expect to find distributions of the same manner. We analyze the frequency distributions for productivity at the field and discipline levels, using the dataset of all Italian professors in the period 2009-2013. The aim of the paper is twofold. First, we intend to provide national and global readers with benchmarks of the yearly average productivity distribution in each field. Next, and more immediately interesting, we wish to investigate the between-field variation of skewness of productivity distributions and their fractal nature. More specifically, we try to answer the following questions:

- Is productivity distribution highly skewed in every field?
- Do the different fields within a discipline maintain similar patterns in productivity distribution?
- Are the distributions of a fractal nature, with the same shape in upper tails?
- Do productivity distributions at the discipline level preserve the shape characteristics of the fields? Are the different disciplines similar?

Throughout the paper we account for the fact that data collection and calculation of the FSS indicator can be difficult for some. For this, we also provide field distributions by number of publications alone (found in the Supplementary Material), and repeat several steps of the analysis using these.

In the next section of the paper we describe our data sources, indicators and the methodology used for the analyses. Sections 3 and 4 present the results and our conclusions.

## 2. Data and method

Data on Italian academics and their SDS classifications are extracted from the database of the Ministry of Education, Universities and Research (MIUR).[2]

Research productivity is measured by FSS, i.e. the yearly total impact of an individual's research activity over a period of time, adopting the fractional counting method.[3]

$$FSS = \frac{1}{t}\sum_{i=1}^{N}\frac{c_i}{\bar{c}}f_i$$

[1]

---

[2] http://cercauniversita.cineca.it/php5/docenti/cerca.php, last accessed on January 23, 2017.

[3] The detailed description of the indicator, as well as the underlying microeconomic theory, can be found in Abramo and D'Angelo (2014).

Where:

$t$ = number of years of work in the period under observation
$N$ = number of publications in the period under observation
$c_i$ = citations received by publication $i$
$\bar{c}$ = average of distribution of citations received for all cited Italian publications[4] in same year and subject category of publication $i$
$f_i$ = fractional contribution of author to publication $i$

The fractional contribution equals the inverse of the number of authors, in those fields where the order of the authors in the byline is mainly alphabetical or without consistent pattern. For the life sciences, where ordering is typically based on personal contribution, we assign different weights according to byline position and the character of co-authorship (intra-mural or extra-mural) (Abramo, D'Angelo, & Rosati, 2013).

The bibliometric dataset is extracted from the Italian Observatory on Public Research (ORP), a database developed and maintained by the authors. The database is derived from the WoS, and consists of all publications (articles, reviews, letters and conference proceedings) produced by researchers in Italian public institutions, since 2001. Based on these data, using a complex algorithm[5] for reconciling author affiliation and disambiguating their true identities, each publication is attributed to the research scientist(s) who produced it.[6]

We construct the dataset beginning with the hard science disciplines, for which bibliometric indicators are generally accepted as effective measures of performance. In the Italian case, scientists are grouped into 205 fields (SDSs), under nine disciplines (UDAs). We also include the Psychology discipline, having verified that the WoS contains adequate coverage of production by Italian academics in this discipline. The observations concern all production from 2009 to 2013, with citations to publications counted as of 31/12/2015. For reasons of robustness, we exclude those SDSs with less than 30 professors, although random fluctuations may still occur for small-sized SDSs. The dataset is thus composed of 31,532 Italian professors, belonging to 180 SDSs,[7] having a total of 230,731 publications. Table 1 provides the breakdown by discipline.

Knowing the assignment of each and every professor to their one specific SDS, along with the observations of their publications, we can investigate the productivity distributions at the field (SDS) level and subsequently proceed to discipline (UDA) level. We analyse the distributions using the Characteristic Scores and Scale (CSS) technique, developed by Schubert, Glänzel, & Braun (1987), and already applied to bibliometric studies (Glänzel & Schubert, 1988: Glänzel, 2011; Ruiz-Castillo & Costas, 2014; Bornmann & Glänzel, in press). The technique involves reiterated truncation of a frequency distribution according to mean values, also called "characteristic scores". After truncating the overall distribution at its mean value, the mean of the subpopulation above the first mean is recalculated; the subpopulation is again truncated, and so on until the procedure is stopped. Applying the CSS method up to three characteristic

---

[4] Abramo, Cicero, & D'Angelo (2012) demonstrate that the most effective scaling factor is provided by the average of citations for all cited Italian publications of the same year and subject category.

[5] D'Angelo, Giuffrida, & Abramo (2011) describes the algorithm in full.

[6] The harmonic mean of precision and recall (F-measure) of authorships, as disambiguated by the algorithm, is around 97% (2% margin of error, 98% confidence interval).

[7] The complete list of SDSs is accessible at:
http://www.iasi.cnr.it/laboratoriortt/TESTI/Indicators/Appendix_Skewness.pdf (last accessed on January 23, 2017).

scores, we obtain the following five categories for each field distribution under examination:
- Unproductive professors (UP): FSS = 0
- Low performers (LP): 0 < FSS ≤ $\mu_1$
- Fair performers (FP): $\mu_1$ < FSS ≤ $\mu_2$
- High performers (HP): $\mu_2$ < FSS ≤ $\mu_3$
- Very high performers (VHP): FSS > $\mu_3$

Where:

$\mu_1$ = mean value of the overall population

$\mu_2$ = mean value of the population above $\mu_1$

$\mu_3$ = mean value of the population above $\mu_2$

*Table 1: Dataset: number of SDSs (with at least 30 professors), professors and publications, per UDA (2009-2013 data)*

| UDA | SDSs | Professors | Publications |
|---|---|---|---|
| 1 - Mathematics and computer science | 10 | 2,893 | 17,590 |
| 2 - Physics | 7 | 1,968 | 25,418 |
| 3 - Chemistry | 11 | 2,646 | 27,344 |
| 4 - Earth sciences | 11 | 946 | 6,506 |
| 5 - Biology | 19 | 4,391 | 36,639 |
| 6 - Medicine | 47 | 9,067 | 78,468 |
| 7 - Agricultural and veterinary sciences | 28 | 2,743 | 15,412 |
| 8 - Civil engineering | 8 | 1,378 | 7,970 |
| 9 - Industrial and information engineering | 31 | 4,442 | 44,390 |
| 11 - Psychology | 8 | 1,058 | 5,429 |
| Total | 180 | 31,532 | 230,731* |

\* *The total is less than the sum of column data due to multiple counting of publications authored by professors of more than one UDA.*

Following the approach of Ruiz-Castillo and Costas (2014), we then measure the degree of SDS skewness for each distribution, using the index proposed by Groeneveld and Meeden (1984), calculated as:

$$GM = \frac{\mu - Q_2}{E|X - Q_2|}$$

[2]

Where:

X = random variable with a continuous distribution function F(x)

$Q_2$ = the second quartile of X (i.e. the median)

$\mu$ = mean of the distribution

E = expected value

The "GM index"[8] is a continuous variable assuming values between +1 and −1, with positives (the mean greater than the median) showing right skewness, and negative values showing left skewness. A zero value suggests symmetric distribution.

For those readers who are interested, we repeat the same analysis by average number of publications (PO) per year. The full results of this second analysis are found in the Supplementary Material to the paper.

---

[8] It is one of the extensions of the Bowley coefficient of skewness, under the assumption that X has a continuous distribution function F(X) with a differentiable density function f(x)>0 on an interval I=(a, b), where *a* can be -∞ or +∞ respectively. The formula's denominator represents the average of the absolute deviations from the data's median. The numerator is a difference between distributional mean and median.

## 3. Results and analysis

The analysis of skewness in productivity distributions is first carried out at the field (SDS) level. Between-field differences are assessed within the respective disciplines (UDAs). We warn the reader against possible random fluctuations in small-sized SDSs. For brevity, the paper shows and discusses the application of CSS technique to only one discipline (Chemistry). The Supplementary Material shows the detailed results for all disciplines. The presentation in the body of the paper is limited to an extract from the results, showing the extreme values of characteristic scores within the 10 disciplines. Finally, we analyze and compare the productivity distributions at the aggregate UDA level.

The skewness of the overall field population (referred to as "A") could be affected by the presence of unproductive and low performers (UP, LP). Therefore we also examine the productivity distribution of the subpopulation above $\mu_1$, roughly corresponding to the upper tail of A. The subpopulation is referred to as "B".

Moreover, a similarity in the partition patterns of these populations might suggest the occurrence of a power law, with property of scale invariance. We refer to such distributions as showing "fractal nature" since, in general, a fractal is an object or pattern that is self-similar across different scales.

### 3.1. The skewness of productivity distributions

For each SDS of Chemistry, Table 2 shows: number of professors; the three characteristic scores of their FSS distributions; the values of GM index for the overall population and the above-$\mu_1$ subpopulation. The results clearly show heterogeneity in the FSS productivity distributions within the Chemistry discipline, in terms of characteristics scores. The differences between SDSs become more remarkable moving towards the upper tails: the coefficients of variation (CVs) for the three characteristic scores are 0.15, 0.17 and 0.30. The GM index values demonstrate that in the majority of cases, productivity of the population above $\mu_1$ has higher degree of skewness, indicating increase of inequality. The opposite occurs in three cases: Foundations of chemistry for technologies (CHIM/07), Food chemistry (CHIM/10), and Chemistry and biotechnology of fermentations (CHIM/11). There is also the interesting situation in which both A and B populations have the same or almost the same GM values: this occurs in Organic chemistry (CHIM/06) and General and inorganic chemistry (CHIM/03).

*Table 2: Characteristic scores and GM indexes for productivity distributions in the SDSs of Chemistry (UDA 3)*

| SDS* | Professors | $\mu_1$ | $\mu_2$ | $\mu_3$ | Skewness Overall population (A) | Skewness Above $\mu_1$ subpopulation (B) |
|---|---|---|---|---|---|---|
| CHIM/01 | 244 | 0.70 | 1.39 | 2.34 | 0.35 | 0.54 |
| CHIM/02 | 363 | 0.71 | 1.62 | 3.25 | 0.53 | 0.65 |
| CHIM/03 | 482 | 0.72 | 1.65 | 2.87 | 0.53 | 0.54 |
| CHIM/04 | 118 | 0.76 | 1.92 | 4.78 | 0.64 | 0.80 |
| CHIM/06 | 540 | 0.62 | 1.38 | 2.40 | 0.56 | 0.56 |
| CHIM/07 | 163 | 0.72 | 1.57 | 2.63 | 0.59 | 0.38 |
| CHIM/08 | 406 | 0.57 | 1.51 | 4.11 | 0.67 | 0.74 |
| CHIM/09 | 183 | 0.54 | 1.05 | 1.66 | 0.27 | 0.57 |
| CHIM/10 | 61 | 0.78 | 1.85 | 2.92 | 0.60 | 0.30 |
| CHIM/11 | 36 | 0.50 | 1.39 | 2.29 | 0.75 | 0.39 |
| CHIM/12 | 50 | 0.56 | 1.21 | 3.06 | 0.53 | 0.87 |

\* CHIM/01, Analytical chemistry; CHIM/02, Physical chemistry; CHIM/03, General and inorganic chemistry; CHIM/04, Industrial chemistry; CHIM/06, Organic chemistry; CHIM/07, Foundations of chemistry for technologies; CHIM/08, Pharmaceutical chemistry; CHIM/09, Applied technological pharmaceutics; CHIM/10, Food chemistry; CHIM/11, Chemistry and biotechnology of fermentations; CHIM/12, Environmental chemistry and chemistry for cultural heritage.

Table 3 shows the partition of SDS distributions into five categories by the CSS approach, reporting the percentage of professors in each category. The differences across SDSs are most pronounced in the extreme categories of productivity. The variability of percentages across disciplines, as measured by CV, reaches maximum value (1.03) in the UP category, and is lowest (0.07) in the LP category. From here it increases towards the opposite end of the productivity spectrum, up to 0.36 in the VHP category. The share of professors in UP category varies from zero in Industrial chemistry (CHIM/04) and Food chemistry (CHIM/10), to 8.3% in Chemistry and biotechnology of fermentations (CHIM/11). The shares in VHP range from 0.5% in Pharmaceutical chemistry (CHIM/08) to a maximum of 5.6% in Chemistry and biotechnology of fermentations (CHIM/11).

*Table 3: FSS distribution partitions (percentage of professors in each category) in the SDSs of Chemistry*

| SDS* | Professors | UP | LP | FP | HP | VHP |
|---|---|---|---|---|---|---|
| CHIM/01 | 244 | 1.6 | 62.7 | 25.0 | 7.0 | 3.7 |
| CHIM/02 | 363 | 1.4 | 67.5 | 23.1 | 5.0 | 3.0 |
| CHIM/03 | 482 | 3.3 | 64.3 | 22.2 | 7.1 | 3.1 |
| CHIM/04 | 118 | 0.0 | 72.9 | 21.2 | 2.5 | 3.4 |
| CHIM/06 | 540 | 1.7 | 66.7 | 21.9 | 6.9 | 3.0 |
| CHIM/07 | 163 | 2.5 | 65.0 | 22.1 | 6.7 | 3.7 |
| CHIM/08 | 406 | 1.2 | 71.2 | 22.2 | 4.9 | 0.5 |
| CHIM/09 | 183 | 1.1 | 58.5 | 27.3 | 9.8 | 3.3 |
| CHIM/10 | 61 | 0.0 | 68.9 | 19.7 | 8.2 | 3.3 |
| CHIM/11 | 36 | 8.3 | 63.9 | 16.7 | 5.6 | 5.6 |
| CHIM/12 | 50 | 4.0 | 58.0 | 30.0 | 4.0 | 4.0 |

\* CHIM/01, Analytical chemistry; CHIM/02, Physical chemistry; CHIM/03, General and inorganic chemistry; CHIM/04, Industrial chemistry; CHIM/06, Organic chemistry; CHIM/07, Foundations of chemistry for technologies; CHIM/08, Pharmaceutical chemistry; CHIM/09, Applied technological pharmaceutics; CHIM/10, Food chemistry; CHIM/11, Chemistry and biotechnology of fermentations; CHIM/12, Environmental chemistry and chemistry for cultural heritage.

To assess the possibility of fractal nature in productivity distributions, we first analyze populations A and B separately and then compare. Each is partitioned into three CSS categories, representing the shares of professors below, above and between characteristic scores: $\mu_1$ and $\mu_2$ scores for population A; $\mu_2$ and $\mu_3$ for B.

For the A populations of the Chemistry SDSs, Figure 1 shows the bar chart of FSS distributions partitioned into UP+LP; FP; and HP+VHP. The two vertical dashed lines represent the mean shares in the first and last category across SDSs. On average, 67.7% of professors have an FSS value below $\mu_1$ and 9.5% above $\mu_2$. The least skewed distribution is observed in Applied technological pharmaceutics (CHIM/09). This is confirmed by its GM index, which is the lowest for all SDSs (Table 2, column 6).

*Figure 1: Partitioning of productivity distributions (FSS) by the CSS technique, for SDSs in Chemistry: overall population (A)*

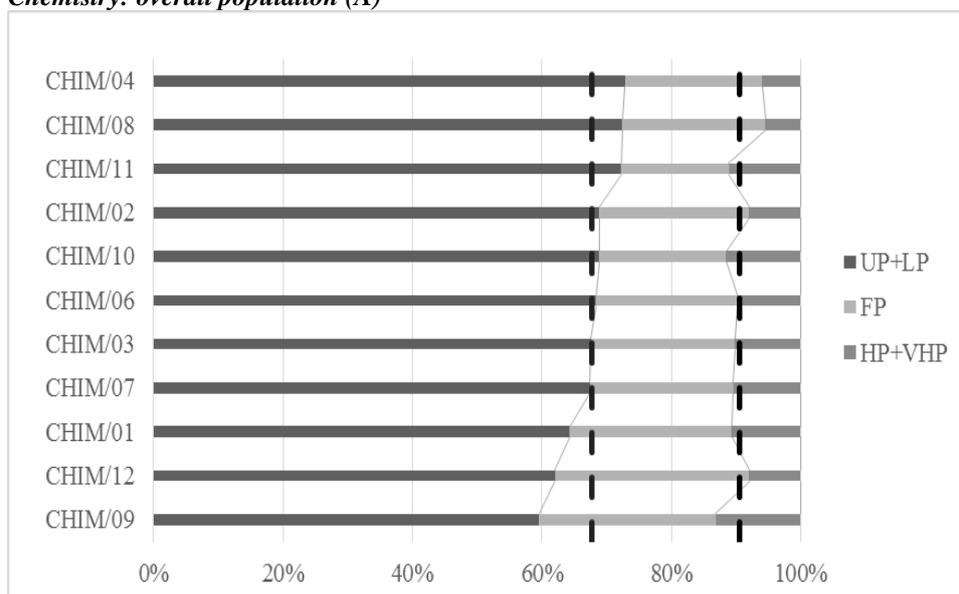

Figure 2 shows the partitioning of the subpopulation above $\mu_1$ (B) into three categories and again presents that for the overall population (A), to facilitate comparison of distribution shapes and assessment of fractal nature. Only three fields demonstrate highly similar patterns of distribution in the two populations: Organic chemistry (CHIM/06), General and inorganic chemistry (CHIM/03), and Foundations of chemistry for technologies (CHIM/07).

*Figure 2: Partitioning of productivity distributions (FSS) by the CSS technique, for SDSs in Chemistry: subpopulation with FSS above $\mu_1$ (B) and overall population (A)*

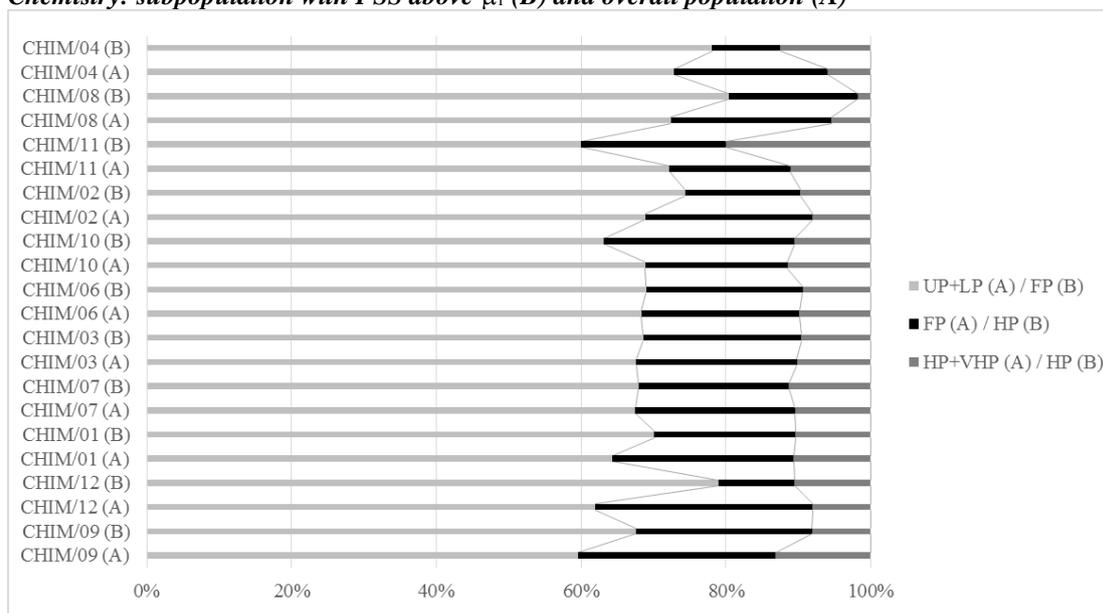

We have repeated the same analysis for all other disciplines (see the Supplementary Material). In the body of the paper we report only the SDSs with the minimum and maximum values of characteristic scores within each UDA (Table 4). The differences are notable not only within disciplines, but also between. The latter is not at all surprising, as the FSS indicator embeds both the number of publications and their individual citations. Therefore, in publication-intensive disciplines such as Chemistry (UDA 3) and Physics (UDA 2) the second and third mean values are much higher than in disciplines known for low publication, such as Psychology (UDA 11) or Civil engineering (UDA 8). We also note that the low value of the first mean ($\mu_1$) is affected by the share of UPs in the overall distribution.

The range of the shares of professors in the five CSS categories in each UDA is shown in Table 5. The maximum of 65.3% in UP category occurs in Complementary mathematics (MAT/04), of the Mathematics and computer science UDA, while the minimum of zero is observed in several SDSs under three UDAs: Aerospace construction and installation (ING-IND/34), in the Industrial and information engineering UDA; Industrial chemistry (CHIM/04) and Food chemistry (CHIM/10) in Chemistry UDA; Veterinary physiology (VET/02), Parasitology and parasitic animal diseases (VET/06) and Veterinary pharmacology and toxicology (VET/07), in the Agricultural and veterinary sciences UDA. The highest and lowest shares of VHPs both occur in the discipline of Industrial and information engineering: a maximum of 8.1% in Mechanical and thermal measuring systems (ING-IND/12); minimum (nil) in the two fields of Aerospace construction and installation (ING-IND/04) and Nuclear plants (ING-IND/19). Nil values indicate that in these SDSs there is one professor who outperforms all the others. Their productivity drives up the mean values, while positioning them alone above $\mu_2$, meaning that their FSS value is itself the third mean and there are no observations above it.

*Table 4: Characteristic scores for discipline productivity distributions: SDSs showing min/max values under each UDA*

| UDA* | $\mu_1$ Min | | $\mu_1$ Max | | $\mu_2$ Min | | $\mu_2$ Max | | $\mu_3$ Min | | $\mu_3$ Max | |
|---|---|---|---|---|---|---|---|---|---|---|---|---|
| 1 | 0.04 | MAT/04 | 0.64 | MAT/08 | 0.20 | MAT/04 | 1.65 | MAT/08 | 0.43 | MAT/04 | 3.48 | MAT/05 |
| 2 | 0.29 | FIS/06 | 0.75 | FIS/03 | 0.64 | FIS/06 | 1.68 | FIS/03 | 0.94 | FIS/06 | 2.91 | FIS/03 |
| 3 | 0.50 | CHIM/11 | 0.78 | CHIM/10 | 1.05 | CHIM/09 | 1.92 | CHIM/04 | 1.66 | CHIM/09 | 4.78 | CHIM/04 |
| 4 | 0.22 | GEO/11 | 0.60 | GEO/03 | 0.51 | GEO/04 | 1.28 | GEO/03 | 0.82 | GEO/04 | 2.22 | GEO/03 |
| 5 | 0.17 | BIO/08 | 0.67 | BIO/12 | 0.42 | BIO/08 | 2.30 | BIO/12 | 0.66 | BIO/08 | 4.90 | BIO/12 |
| 6 | 0.06 | MED/02 | 0.98 | MED/16 | 0.25 | MED/02 | 2.60 | MED/11 | 0.34 | MED/02 | 4.93 | MED/11 |
| 7 | 0.08 | AGR/01 | 0.75 | VET/06 | 0.27 | VET/09 | 2.18 | VET/06 | 0.43 | VET/09 | 4.92 | VET/06 |
| 8 | 0.12 | ICAR/06 | 0.60 | ICAR/08 | 0.35 | ICAR/06 | 1.45 | ICAR/08 | 0.62 | ICAR/06 | 2.70 | ICAR/08 |
| 9 | 0.17 | ING-IND/15 | 0.94 | ING-IND/34 | 0.49 | ING-IND/12 | 3.23 | ING-IND/19 | 0.70 | ING-IND/12 | 15.56 | ING-IND/04 |
| 11 | 0.09 | M-PSI/07 | 0.93 | M-PSI/02 | 0.35 | M-PSI/07 | 1.87 | M-PSI/02 | 0.73 | M-PSI/07 | 2.92 | M-PSI/02 |

\* 1, Mathematics and computer science; 2, Physics; 3, Chemistry; 4, Earth sciences; 5, Biology; 6, Medicine; 7, Agricultural and veterinary sciences; 8, Civil engineering; 9, Industrial and information engineering; 11, Psychology

*Table 5: Partitioning of discipline productivity distributions: fields (SDSs) with min/max shares, by percentage of professors, in each UDA*

| UDA* | | UP | | LP | | FP | | HP | | VHP | |
|---|---|---|---|---|---|---|---|---|---|---|---|
| 1 | min | 4.0 | MAT/09 | 18.1 | MAT/04 | 11.1 | MAT/04 | 4.2 | MAT/04 | 1.4 | MAT/04 |
| | max | 65.3 | MAT/04 | 59.6 | MAT/08 | 26.4 | MAT/09 | 10.4 | MAT/06 | 5.6 | MAT/09 |
| 2 | min | 3.5 | FIS/03 | 57.1 | FIS/02 | 21.1 | FIS/05 | 6.5 | FIS/07 | 2.2 | FIS/07 |
| | max | 7.1 | FIS/02 | 64.5 | FIS/05 | 24.2 | FIS/01 | 11.6 | FIS/04 | 7.0 | FIS/06 |
| 3 | min | 0.0 | CHIM/04;10 | 58.0 | CHIM/12 | 16.7 | CHIM/11 | 2.5 | CHIM/04 | 0.5 | CHIM/08 |
| | max | 8.3 | CHIM/11 | 72.9 | CHIM/04 | 30 | CHIM/12 | 9.8 | CHIM/09 | 5.6 | CHIM/11 |
| 4 | min | 2.9 | GEO/08 | 53.1 | GEO/05 | 19.5 | GEO/05 | 2.4 | GEO/11 | 1.8 | GEO/05 |
| | max | 19.5 | GEO/05 | 66.7 | GEO/11 | 26.7 | GEO/01 | 9.7 | GEO/09 | 7.2 | GEO/07 |
| 5 | min | 1.3 | BIO/15 | 49.1 | BIO/08 | 15.8 | BIO/12 | 4.1 | BIO/04 | 1.3 | BIO/16 |
| | max | 18.9 | BIO/08 | 74.0 | BIO/12 | 24 | BIO/03 | 9.0 | BIO/15 | 5.2 | BIO/04 |
| 6 | min | 2.2 | MED/08 | 25.8 | MED/02 | 9.7 | MED/02 | 3.0 | MED/14 | 1.0 | MED/04 |
| | max | 51.6 | MED/02 | 73.1 | MED/01 | 25.5 | MED/46 | 9.7 | MED/02 | 7.9 | MED/37 |
| 7 | min | 0.0 | VET/02;06;07 | 30.6 | AGR/01 | 8.7 | AGR/17 | 1.0 | VET/05 | 1.6 | VET/06 |
| | max | 45.8 | AGR/01 | 82.6 | AGR/17 | 30.1 | AGR/07 | 11.8 | VET/08 | 7.9 | VET/04 |
| 8 | min | 5.8 | ICAR/03 | 34.6 | ICAR/06 | 19.6 | ICAR/06 | 4.7 | ICAR/03 | 2.3 | ICAR/03 |
| | max | 35.5 | ICAR/06 | 66.3 | ICAR/03 | 25 | ICAR/05 | 7.6 | ICAR/01 | 6.1 | ICAR/07 |
| 9 | min | 0.0 | ING-IND/34 | 49.1 | ING-IND/11 | 14.6 | ING-IND/15 | 2.0 | ING-IND/04 | 0.0 | ING-IND/04;19 |
| | max | 23.2 | ING-IND/15 | 68.6 | ING-IND/04 | 26.9 | ING-IND/31 | 11.1 | ING-INF/07 | 8.1 | ING-IND/12 |
| 11 | min | 0.9 | M-PSI/02 | 21.4 | M-PSI/07 | 15.5 | M-PSI/07 | 2.8 | M-PSI/06 | 1.9 | M-PSI/08 |
| | max | 56.3 | M-PSI/07 | 63.0 | M-PSI/02 | 23.1 | M-PSI/02 | 12.3 | M-PSI/03 | 4.7 | M-PSI/04 |

*\* 1, Mathematics and computer science; 2, Physics; 3, Chemistry; 4, Earth sciences; 5, Biology; 6, Medicine; 7, Agricultural and veterinary sciences; 8, Civil engineering; 9, Industrial and information engineering; 11, Psychology*

## 3.2. The fractal nature of productivity distributions

We now assess the fractal nature of the productivity distributions in each SDS. We proceed as follows. For each population we measure the ratio of the share of professors falling into contiguous categories, whereby the first contiguous category is the denominator. Since we have three categories in each population, we end up with two ratios. Because productivity distributions are generally right skewed, the value of the ratio is in most cases below 1. We then name the ratios "decay ratio" or DR. We then calculate the (absolute value) differences between the corresponding DR values of populations A and B. We refer to the first difference as to DDR1, and to the second as DDR2. The closer the DDR value is to zero, the stronger the evidence of the fractal nature of a distribution, as it indicates that the shares of professors in contiguous categories of the two populations increase/decrease similarly.

Let us consider the example of Applied geophysics (GEO/11). Table 6 presents the partitioning of populations A and B into three categories, along with the values of DR and DDR. In population A, the ratio of shares of professors of FP category to UP+LP leads to a DR1(A) value of 0.3, while in subpopulation B the ratio DR1(B) of the first two partition categories HP and FP equals 0.1. The same logic is applied for calculation of DR2(A) and DR2(B). The DDR1 (0.2) is the absolute difference between DR1(A) (0.3) and DR1(B) (0.1). The value of DDR2 (1.7) reveals a substantial difference in the tails of the two population distributions: in population A, DR2(A) equals 0.3, while in subpopulation B the value 2.0 for DR2(B) indicates that the share of professors in VHP category exceeds twice the share in HP. Comparing the values of DDR1 and DDR2, we can reject the fractal nature of the productivity distribution in this SDS.

*Table 6: CSS partition, DR and DDR values of populations A and B in GEO/11-Applied geophysics*

| Overall population (A) | | | Above $\mu_1$ subpopulation (B) | | |
|---|---|---|---|---|---|
| UP+LP | FP | HP+VHP | FP | HP | VHP |
| 71.4% | 21.4% | 7.1% | 75.0% | 8.3% | 16.7% |
| DR1(A) | | DR2(A) | DR1(B) | | DR2(B) |
| 21.4/71.4=0.3 | | 7.1/21.4=0.3 | 8.3/75.0=0.1 | | 16.7/8.3=2.0 |
| DDR1 | | | DDR2 | | |
| \|0.3-0.1\|=0.2 | | | \|0.3-2.0\|=1.7 | | |

We repeat the same steps for all SDSs, and present a scatter plot of DDR1 against DDR2 in Figure 3. The results are vastly dispersed, with the three most distant points having values 0.2 and 2.8 in Infectious diseases of domestic animals (VET/05); 0.2 and 0.7 in Applied geophysics (GEO/11); and 0.9 and 1.0 in History of medicine (MED/02). The productivity distributions of the SDSs close to the origin are likely to have a fractal nature. To delve into these particular distributions, we divide the Cartesian plane into four sections by the median values of each axis, and then enlarge the scale of the section including the origin in the upper right corner. Fifty-six (31%) out of 180 SDSs fall in that section. The points are evenly dispersed and do not seem to show any particular dependencies. To check for the robustness of results, we have repeated the same analysis for the larger SDSs only, i.e. those falling in the first quartile by size (corresponding to a threshold of 197 professors), 45 in all. In this case, we found 11 (24%) SDSs, falling in the section including the origin.

*Figure 3: Scatter plot of DDR1 against DDR2 for all SDSs*

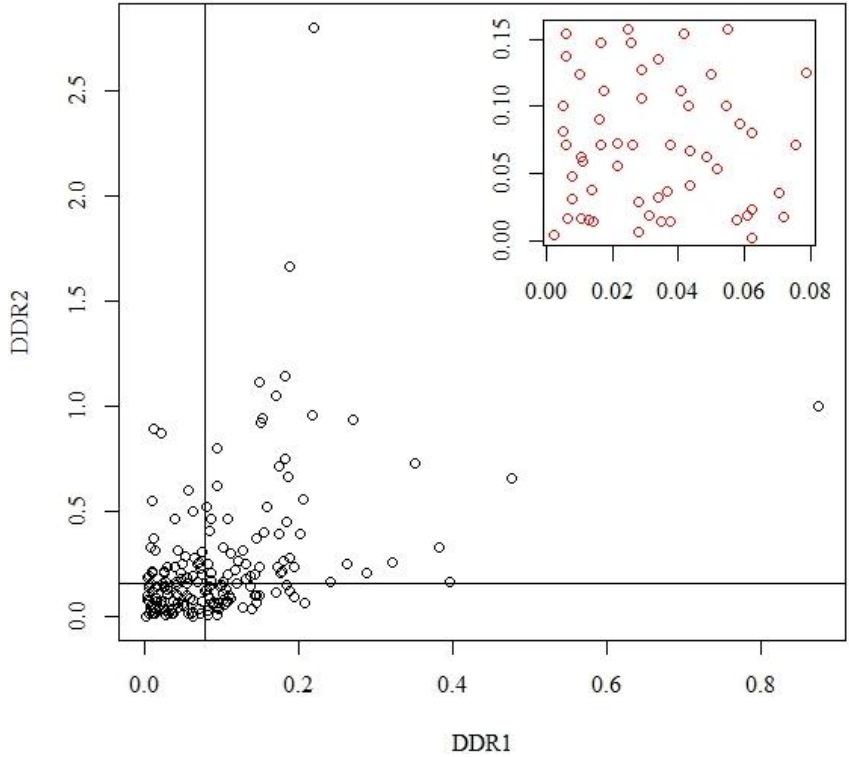

For each UDA, Table 7 presents the SDSs with the min/max GM index along with the coefficient of variation (CV) for both populations A and B. As for population A, the GM index reaches its upper bound (1) in three SDSs: Complementary mathematics (MAT/04), History of medicine (MED/02) and Dynamic psychology (M-PSI/07). This is due to the share of UP, which is above 50% (see figures in Table SM2 of the Supplementary Material), and thus causes the FSS median to equal nil. The minimum values of GM range from 0.25 in Geochemistry and volcanology (GEO/08) to 0.51 in Nephrology (MED/14). The lowest variability in the degree of skewness, measured by CV (0.13), occurs in Biology (UDA 5), while the highest one (0.33) occurs in Industrial and information engineering (UDA 9).

As for population B, the maximum GM ranges from 0.67 in Maritime hydraulic construction and hydrology (ICAR/02) to 0.92 in Nuclear Plants (ING-IND/19). The minimum values range from 0.06 in Mechanical and thermal measuring systems (ING-IND/12) to 0.47 in Probability and mathematical statistics (MAT/06). The lowest variability in degree of skewness measured by CV (0.15) is observed in Mathematics and computer science (UDA 1), while the highest one (0.80) is in Industrial and information engineering (UDA 9), as in the case of population A.

*Table 7: SDSs with min/max degrees of skewness (GM index) and coefficient of variation (CV) in each UDA*

| UDA* | Overall population (A) | | | | | Above $\mu_1$ subpopulation (B) | | | | |
|---|---|---|---|---|---|---|---|---|---|---|
| | GM_Min | | GM_Max | | CV | GM_Min | | GM_Max | | CV |
| 1 | 0.41 | MAT/09 | 1.00 | MAT/04 | 0.26 | 0.47 | MAT/06 | 0.77 | MAT/04 | 0.15 |
| 2 | 0.49 | FIS/02 | 0.66 | FIS/06 | 0.29 | 0.24 | FIS/04 | 0.70 | FIS/07 | 0.35 |
| 3 | 0.27 | CHIM/09 | 0.75 | CHIM/11 | 0.25 | 0.30 | CHIM/10 | 0.87 | CHIM/12 | 0.31 |
| 4 | 0.25 | GEO/08 | 0.75 | GEO/05 | 0.27 | 0.19 | GEO/09 | 0.73 | GEO/05 | 0.42 |
| 5 | 0.50 | BIO/07;BIO/05 | 0.78 | BIO/12 | 0.13 | 0.35 | BIO/19 | 0.81 | BIO/03 | 0.21 |
| 6 | 0.51 | MED/14 | 1.00 | MED/02 | 0.15 | 0.31 | MED/23 | 0.80 | MED/10 | 0.42 |
| 7 | 0.30 | VET/02 | 0.91 | AGR/01 | 0.24 | 0.19 | AGR/10 | 0.88 | AGR/04 | 0.34 |
| 8 | 0.48 | ICAR/01 | 0.85 | ICAR/06 | 0.20 | 0.36 | ICAR/07 | 0.67 | ICAR/02 | 0.17 |
| 9 | 0.27 | ING-INF/07 | 0.88 | ING-IND/19 | 0.33 | 0.06 | ING-IND/12 | 0.92 | ING-IND/19 | 0.80 |
| 11 | 0.36 | M-PSI/02 | 1.00 | M-PSI/07 | 0.27 | 0.21 | M-PSI/03 | 0.69 | M-PSI/06 | 0.29 |

\* *1, Mathematics and computer science; 2, Physics; 3, Chemistry; 4, Earth sciences; 5, Biology; 6, Medicine; 7, Agricultural and veterinary sciences; 8, Civil engineering; 9, Industrial and information engineering; 11, Psychology*

Figure 4 shows the histograms and boxplots of the GM index in all SDSs for populations A and B. The two distributions are quite similar and their histograms show the bell-shaped pattern. The Shapiro-Wilk normality test gives p-values of 0.138 for population A and 0.077 for population B, consequently we cannot reject the hypothesis of normality at the 0.95 confidence level. Then we can use the mean value as an appropriate metric to describe the degree of skewness in the SDSs. On average, productivity distributions result highly skewed for both populations A and B, with the highest frequency concentrated between 0.5 and 0.7, and respective mean values of 0.61 and 0.55.

Both boxplots of the GM distributions show symmetrical shapes with the medians roughly in the middle of the box, representing values of 0.60 for population A, and 0.56 for population B. The height of the box is represented by interquartile range (IQR), showing the spread of values: 0.18 in both populations. The inter-whiskers range is 0.64 for population A and 0.72 for population B. The outliers, meaning observations more than 1.5 times the IQR beyond either end of the box, all show GM = 1 for the A population. In population B, four SDSs present FSS distributions close to symmetrical, with GM indices of 0.06 in Mechanical and thermal measuring systems (ING-IND/12), 0.14 in Theory of development for chemical processes (ING-IND/26), 0.18 in Metallurgy (ING-IND/21), and 0.19 in Rural construction and environmental land management (AGR/10). The latter findings could be affected by the small number of professors in these SDSs, particularly for population B (see Table SM3 of the Supplementary Material), but in general the correlation between the number of professors per SDS (population B) and the GM index is negligible (Pearson correlation coefficient is equal to 0.157).

*Figure 4: Histograms and boxplots of GM distribution for SDSs in the dataset: overall population (A) and subpopulation with FSS above μ₁ (B)*

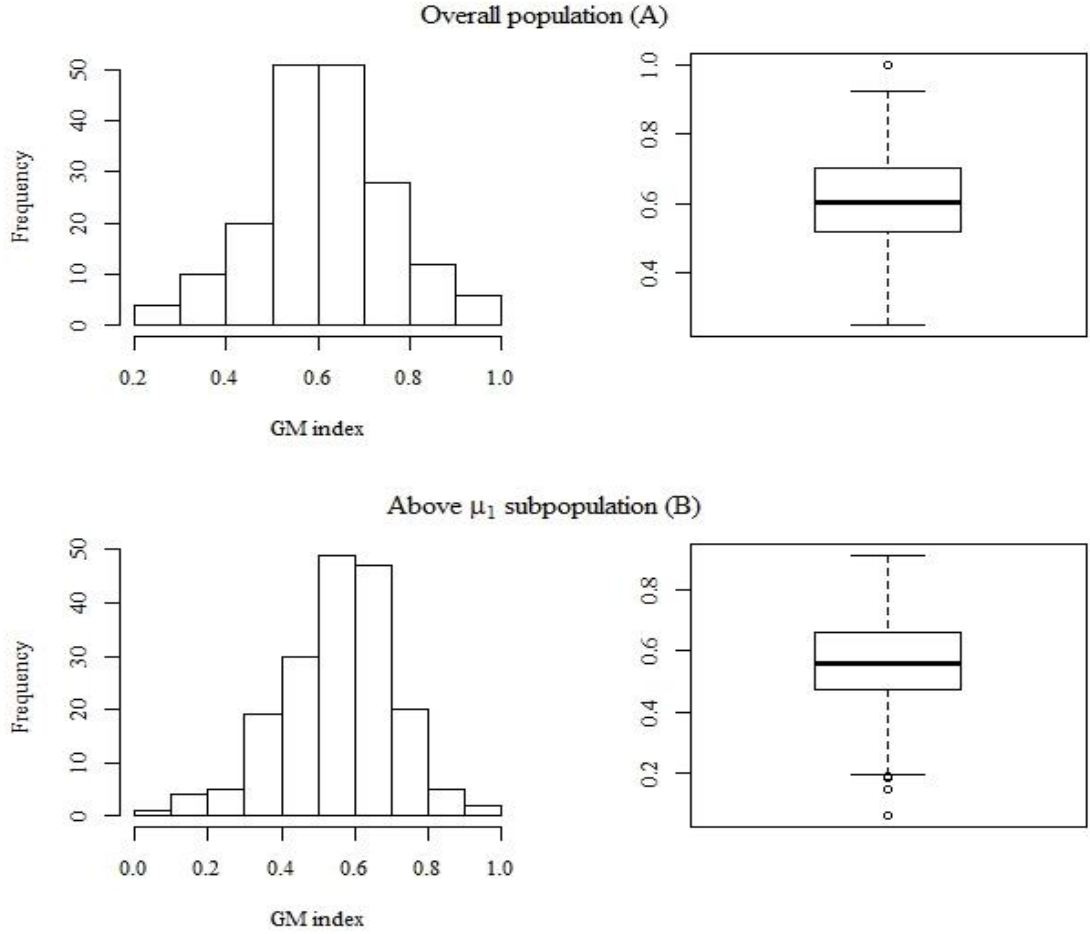

We now turn to the question of similarities in skewness patterns at a higher aggregation level. Table 8 shows the partition of FSS distribution into three categories and the relative shares of professors for both populations A and B at UDA level. The percentages are obtained by counting the professors in the corresponding category in each SDS, summing them up within their corresponding UDAs and calculating the ratio out of the total amount present in each UDA. The shares in the last row of Table 8, which embeds all UDAs, are obtained likewise.

The results show values not drastically different across UDAs for each CSS category. The comparison of population A and B reveals very similar partition patterns for distributions in Mathematics and computer science (UDA 1), Earth sciences (UDA 4), Industrial and information engineering (UDA 9) as well as in total, implying a fractal nature at least at this aggregate level of analysis.

*Table 8: Comparison of FSS distribution partitions (percentage of professors in each category) at UDA level*

| UDA* | Overall population (A) | | | Above $\mu_1$ subpopulation (B) | | |
|---|---|---|---|---|---|---|
| | UP+LP | FP | HP+VHP | FP | HP | VHP |
| 1 - Mathematics and computer science | 69.9 | 21.0 | 9.1 | 69.8 | 21.8 | 8.4 |
| 2 - Physics | 66.3 | 23.1 | 10.6 | 68.5 | 23.0 | 8.5 |
| 3 - Chemistry | 68.0 | 22.8 | 9.2 | 71.4 | 19.7 | 8.9 |
| 4 - Earth sciences | 66.2 | 22.6 | 11.2 | 66.9 | 20.6 | 12.5 |
| 5 - Biology | 70.7 | 20.1 | 9.2 | 68.7 | 20.8 | 10.5 |
| 6 - Medicine | 72.5 | 19.0 | 8.5 | 69.1 | 20.1 | 10.8 |
| 7 - Agricultural and veterinary sciences | 68.1 | 20.9 | 11.0 | 65.5 | 20.8 | 13.7 |
| 8 - Civil engineering | 66.6 | 22.5 | 10.9 | 67.4 | 20.0 | 12.6 |
| 9 - Industrial and information engineering | 67.2 | 21.9 | 10.9 | 66.8 | 21.8 | 11.4 |
| 11 - Psychology | 72.0 | 19.1 | 8.9 | 68.2 | 20.3 | 11.5 |
| Total | 69.7 | 20.8 | 9.6 | 68.4 | 20.9 | 10.7 |

To investigate the fractal nature of productivity distributions at the discipline level, we repeat the analysis carried out at the field level in Chemistry (Figure 2). Figure 5 illustrates the UDA distribution shapes for populations A and B. Notwithstanding the notable differences between SDSs within each UDA observed above (Table 7), distribution patterns at aggregate level are much more similar and highly skewed. For population A, the average value by UDA of percentages of professors with FSS below $\mu_1$ is 68.7%, while for those above $\mu_2$ the average is 9.9%. For population B, the average value by UDA of percentages of professors with FSS below $\mu_2$ is 68.2%, while for those above $\mu_3$ it is 10.8%.

*Figure 5: FSS distribution partitions according to the CSS technique at UDA level: subpopulation with FSS above $\mu_1$ (B) and overall population (A)*

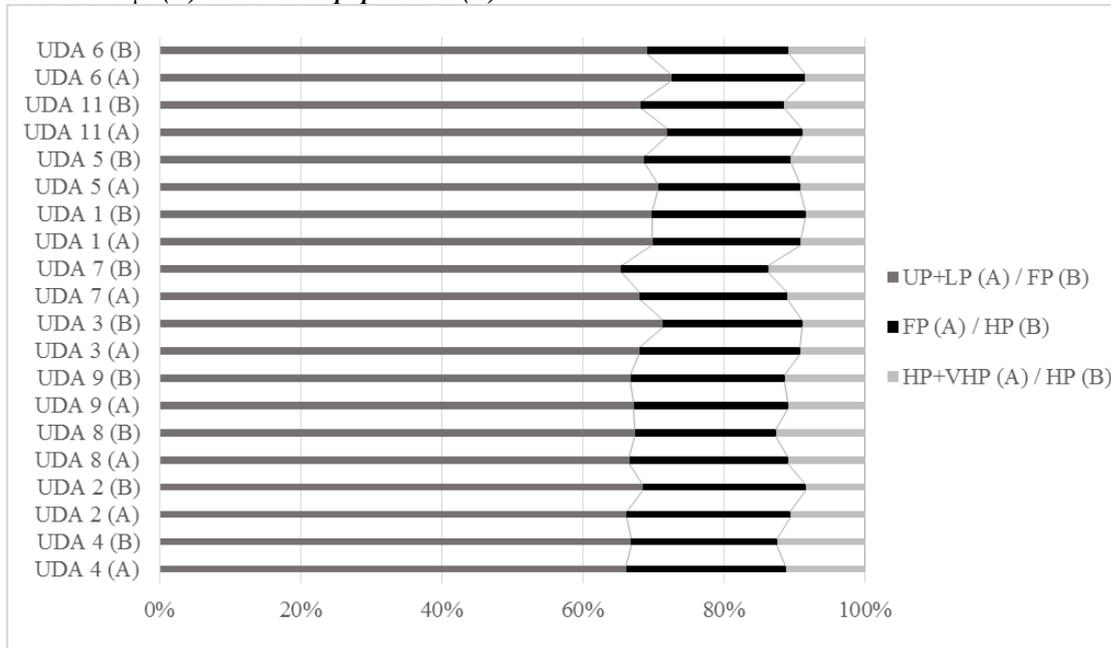

*1 - Mathematics and computer science; 2 - Physics; 3 - Chemistry; 4 - Earth sciences; 5 - Biology; 6 - Medicine; 7 - Agricultural and veterinary sciences; 8 - Civil engineering; 9 - Industrial and information engineering; 11 – Psychology*

For the benefits of those who encounter problems in measuring FSS, we also carry out the analysis of skewness by yearly average output. The full results can be found in the Supplementary Material. Here we provide only a synoptic table. Differently from Ruiz-Castillo and Costas (2014), our results indicate that the shapes of field distributions analyzed by the CSS approach are not very similar. The variability in shares of professors in each category, measured by CV and standard deviations, is higher in subpopulation B (Table 9). The average percentage of professors in the first partition category is 2.1 points higher in subpopulation B, while in the second and third categories the percentages are lower than in population A by 1.3 and 0.9 points.

*Table 9: Yearly average output distributions: variation of share of professors (%) in each partition category across fields: coefficient of variation, mean and standard deviation. Overall population (A) and subpopulation above $\mu_1$ (B)*

|  | Overall population (A) | | | Above $\mu_1$ subpopulation (B) | | |
| --- | --- | --- | --- | --- | --- | --- |
|  | UP + LP | FP | HP + VHP | FP | HP | VHP |
| Average (Std dev) | 62.7 (5.3) | 24.2 (4.0) | 13.2 (3.4) | 64.8 (7.6) | 22.9 (6.7) | 12.3 (4.7) |
| CV | 0.08 | 0.17 | 0.26 | 0.12 | 0.29 | 0.39 |

The histograms and boxplots of GM indicator for the two population groups are presented in Figure 6. Median and mean values are extremely close: 0.40 and 0.39 in population A, and 0.51 and 0.48 in subpopulation B. This indicates a higher degree of skewness in subpopulation B, in accordance with the results presented in Table 9. The inter-whiskers range is 0.64 for population A and 0.65 for B.

In population A the most frequent degree of skewness is between 0.3 and 0.5; the distribution is slightly left skewed and has four outliers. The lowest value of -0.002, indicating symmetrical distribution, is found in Palaeontology and palaeoecology (GEO/01), while the three with top skewness present GM indexes of 0.76 in Experimental physics (FIS/01), 0.78 in Nuclear and subnuclear physics (FIS/04) and 0.81 in History of medicine (MED/02).

The histogram and boxplot of subpopulation B suggest normal distribution in the inter-whiskers range of values between 0.16 and 0.81, and at the same time a large number of outliers. The two extreme observations of -0.69 and -0.49 pertain to Nuclear and subnuclear physics (FIS/04) and Experimental physics (FIS/01). The seven SDSs with approximately symmetrical distributions have GM indexes of -0.02 in Neuroradiology (MED/37); 0.04 in Physical geography and geomorphology (GEO/04), Audiology (MED/32) and Psychology of work and organizations (M-PSI/06); 0.1 in General pathology and veterinary pathological anatomy (VET/03); 0.13 in Clinical veterinary surgery (VET/09) and Geotechnics (ICAR/07). The lone SDS with value exceeding the upper whisker is Dynamic psychology (M-PSI/07), for which the GM index is 0.85. Thus, our findings contrast with those of Ruiz-Castillo and Costas (2014), who reported similarities across fields and lower variability in the population above $\mu_1$.

*Figure 6: Histograms and boxplots of GM distribution for SDSs in the dataset, for yearly average output: overall population and subpopulation above μ₁ (B)*

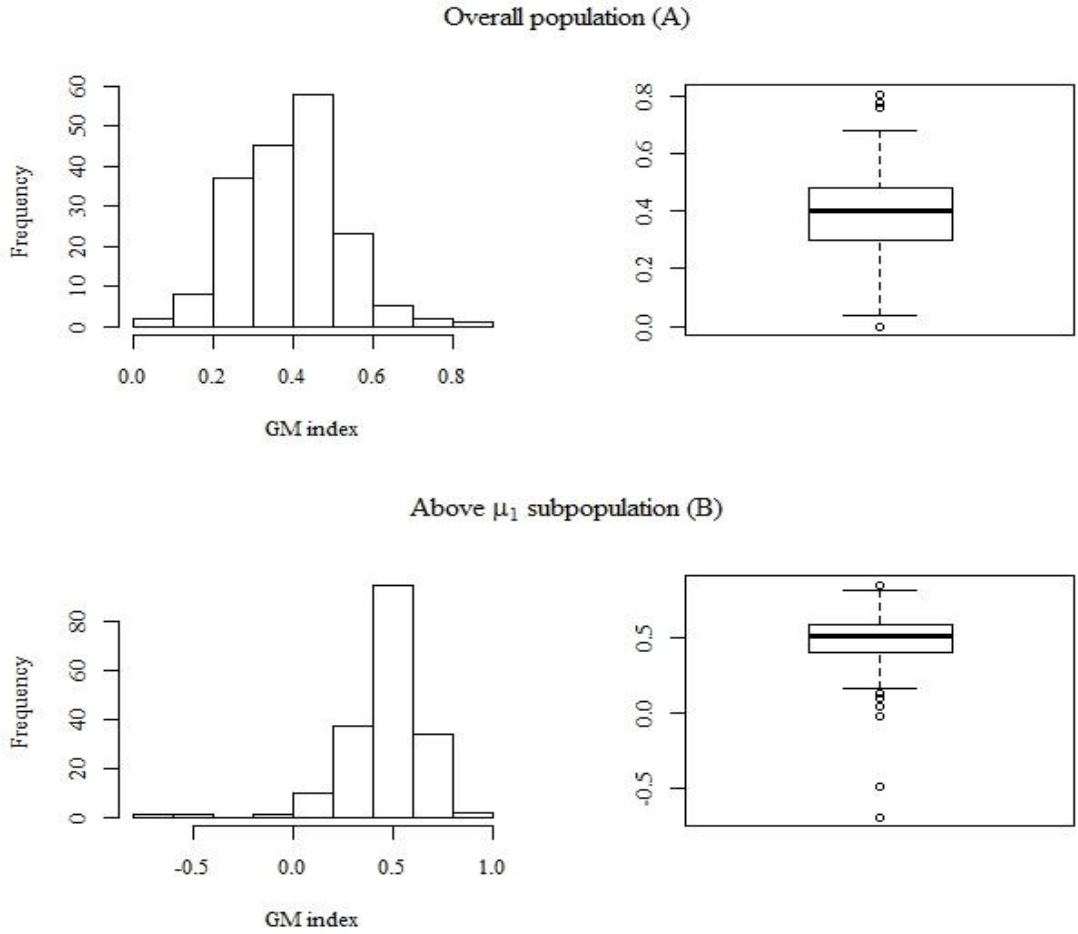

## 4. Conclusions

The study of frequency distributions is an essential step towards evaluation of regularities and trends for any phenomena. In assessment of research it allows better interpretation of results and execution of equitable comparisons between individuals and research units. Inequality in research productivity has attracted attention as one of the most widespread characteristics of scientific activity. As stated by Seglen (1992), inequality is a feature of any highly specialized human activity, always "likely to form an extreme-property distribution".

The present work examines the shape patterns and skewness of productivity distribution at the level of individual scientific field. Results obtained for the population of Italian professors show that, on average, research productivity distribution for all fields is highly skewed to the right, both at overall level and within the upper tail. However, the wide range of GM index values suggests substantial differences in the degree of skewness, and therefore inequality. Identical conclusions are drawn by analyzing distributional partition in categories using the CSS approach. For both FSS and average yearly output, our results are not in line with Ruiz-Castillo and Costas (2014), who found productivity distributional shapes to be very similar across fields.

The reasons could be various. We refer to Italian data only; we use a different productivity indicator, the FSS; we also observe unproductive academics; to measure the productivity of academics we consider the field they belong to and count all their production independently of the subject category where their output falls; and last but not least our field classification is fine-grained (180 fields) as compared to that of the above authors (30 fields). At the same time the results aggregated at discipline level and total reveal analogous highly skewed partition patterns in both populations: overall and above $\mu_1$.

Moreover, our study offers a benchmark of the yearly average productivity distribution shape at the level of individual scientific fields. Given the verification of skewness in the majority of cases, more attention should be paid to the observations in the upper tail as well as to the unproductive category, which is found to have substantial frequency. In addition to all limitations that generally apply to bibliometric studies, we warn the reader against generalization to other national contexts of our findings, which refer to the Italian case.

**Acknowledgments**
The authors are grateful to Javier Ruiz-Castillo for inspiring us to undertake this research work and for extremely helpful comments.